# Diagnosis-Driven Co-planning of Network Reinforcement and BESS for Distribution Grid with High Penetration of Electric Vehicles

Linhan Fang, *Student Member, IEEE,* Elias Raffoul, and Xingpeng Li, *Senior Member, IEEE*

***Abstract*—While the rapid proliferation of electric vehicles (EVs) accelerates net-zero goals, uncoordinated charging activities impose severe operational challenges on distribution grids, including exacerbated peak loads, thermal overloading, and voltage violations. To overcome the computational intractability of jointly optimizing grid infrastructure reinforcements and Battery Energy Storage System (BESS) installations, this paper proposes a novel three-stage Diagnosis-Driven Co-Planning (DDCP) framework. The methodology integrates a Violation Detection and Quantification (VDQ) model to systematically identify system breaches, and a Violation Mitigation-Based Planning (VMBP) model for optimal BESS allocation. Specifically, Stage I of the DDCP framework diagnoses critical bottleneck lines that render standalone BESS solutions infeasible; Stage II executes targeted physical upgrades exclusively on these bottlenecks; and Stage III finalizes the optimal BESS deployment on the updated network topology. Furthermore, this study quantifies the EV hosting capacity thresholds before and after BESS integration across varying EV adoption rates and base voltages. Finally, a comprehensive comparative analysis evaluates four mitigation approaches: the VDQ-driven cable upgrade (VCU) model, the VMBP model, system-wide voltage uprating, and the proposed DDCP framework. The results demonstrate that the DDCP framework not only resolves the complex joint-optimization hurdle but also achieves superior techno-economic performance in addressing high-EV-penetration challenges.**

## NOMENCLATURE

*A. Indices and Sets*

| | |
|---|---|
| $s \in G$ | Index and set of substation bus |
| $i \in N$ | Index and set of non-substation buses |
| $(i,j) \in L$ | Index and set of branch information |
| $t \in T$ | Index and set of time intervals |
| $j \in D(i)$ | Index and set of downstream nodes of node $i$ |
| $k \in U(i)$ | Index and set of upstream nodes of node $i$ |
| $b \in B$ | Index and set of candidate BESS installation buses |

*B. Parameters*

| | |
|---|---|
| $r_{ki}$ | Resistance of the line connecting nodes $k$ and $i$ |
| $x_{ki}$ | Reactance of the line connecting nodes $k$ and $i$ |
| $P_{load,i}^{t}$ | Active power demand of node $i$ at time $t$ |
| $Q_{load,i}^{t}$ | Reactive power demand of node $i$ at time $t$ |
| $E_{min}^{cap}, E_{max}^{cap}$ | Minimum / maximum invested capacity of BESS |
| $SOC^{min}$ | Minimum limit of state of charge of BESS |
| $SOC^{max}$ | Maximum limit of state of charge of BESS |
| $\eta_{charge}$ | Charging power efficiency |
| $\eta_{discharge}$ | Discharging power efficiency |
| $c^{cap}$ | Capacity cost of BESS |
| $C_{rate}^{ch}, C_{rate}^{dis}$ | Charge / discharge C rate of BESS |
| $K_{inv}$ | BESS inverter apparent power sizing factor |

*C. Variables*

| | |
|---|---|
| $P_{ij}^{t}, Q_{ij}^{t}$ | Power flow from node $i$ to node $j$ at time $t$ |
| $P_{s}^{t}, Q_{s}^{t}$ | Total power consumption of the system at time $t$ |
| $v_{i}^{t}$ | The square of voltage magnitude of node $i$ at $t$ |
| $v_{s}^{t}$ | The square of voltage magnitude of bus $s$ at $t$ |
| $V_{s}^{t}$ | Time-varying voltage at substation bus at $t$ |
| $l_{ij}^{t}$ | The square of per-unit current in the line at time $t$ |
| $P_{b,t}^{BESS_ch}$ | Active charging power of BESS at bus $b$ at $t$ |
| $P_{b,t}^{BESS_dis}$ | Active discharging power of BESS at bus $b$ at $t$ |
| $Q_{b,t}^{BESS_inj}$ | Reactive power flowing from BESS into grid at bus $b$ at time $t$ |
| $Q_{b,t}^{BESS_abs}$ | Reactive power flowing from grid into BESS at bus $b$ at time $t$ |
| $E_{b}^{cap}$ | The capacity of BESS at bus $b$ |
| $E_{b,t}^{bess}$ | The amount of the stored energy of BESS of the bus $b$ at time $t$ |
| $S_{inv,b}$ | Inverter capacity for BESS at bus $b$ |

## I. INTRODUCTION

Electric vehicles (EVs) are emerging as a key solution to reduce greenhouse gas emissions from the transportation sector. According to the International Energy Agency's (IEA) "Net Zero Emissions by 2050 Scenario" report, the shift to electric mobility is essential for decarbonizing transportation, which remains a major driver of global emissions [1]. In the US, vehicles are major sources of $CO_2$ emissions, with transportation responsible for about 29% of the nation's greenhouse gas output. Within this sector, passenger cars and light-duty vehicles account for around 58% of emissions [2]-[3].

This growing demand for lower-emission alternatives is reflected in global EV sales, which are hitting record highs. BloombergNEF reports that 14% of new vehicle sales in 2023 were electric—double the share in 2021. Projections suggest EVs could comprise 35-40% of global car sales by 2030, potentially reaching 73% by 2040 [4]. Notably, if the largest automakers meet their targets, IEA predicts over 40 million electric cars could be sold annually by 2030, amplifying the need for infrastructure readiness. Policy support will be critical to accelerating EV adoption, particularly in addressing

Linhan Fang, Elias Raffoul and Xingpeng Li are with the Department of Electrical and Computer Engineering, University of Houston, TX, 77204, USA. (e-mail: lfang7@uh.edu; ejraffou@uh.edu; xli83@central.uh.edu).

challenges related to battery supply chains, and charging networks as EV technology continues to mature [5].

While the rise of EVs offers a promising path to reducing carbon emissions, it also presents challenges to distribution grids. The surge in EV adoption reshapes household energy demands and complicates grid management, especially during evening and nighttime charging hours. Electricity demand from road transport is projected to reach 8.3 petawatt hours by 2050 in the Net Zero Scenario [4]-[5], creating operational challenges such as greater voltage drops, increased power losses, and the risk of overloads, all of which can compromise grid reliability [6]. Studies suggest that integrating controlled charging strategies, particularly those that account for grid constraints, can alleviate strain by distributing peak loads more evenly, enhancing grid reliability [7].

While research provides valuable insights into voltage-related impacts, most studies focus on voltage drops and power quality issues, with fewer addressing line violations or ampacity limits key indicators of grid stress that affect infrastructure longevity and safety. Some probabilistic studies assess the impact of high EV penetration on transformer loading and voltage levels, considering variables like charger ratings and user behaviors, but often overlook line loading violations [8]-[9]. Additionally, many studies focus on low-voltage power systems, neglecting the response of medium-voltage distribution networks, common in suburban areas, under varying voltage conditions [10]. Moreover, existing studies modeling EV penetration levels often use limited ranges, focusing on fixed or narrow adoption rates rather than exploring scenarios from 0% to 100% penetration [11]. Similarly, very few studies conduct sensitivity analyses on varying charging capacities to understand their impact on grid performance under different loading conditions [12]. These gaps highlight the need for more comprehensive approaches that examine a wider range of EV adoption scenarios and charger capacities.

In modern power distribution networks, the escalating demand for EV charging and the rising penetration of distributed energy resources (DERs) have precipitated critical operational challenges [13]-[14]. Specifically, line thermal overloads and nodal voltage violations have emerged as primary constraints on system safety and stability. Conventionally, these issues are addressed through physical-layer network reconfiguration and the adjustment of reactive power compensation devices. These established methods have proven effective for congestion management and voltage support [15]-[16]. However, recent advancements in power electronics have shifted research focus toward volt-var optimization strategies utilizing soft open points and smart inverters. Unlike traditional assets, these technologies offer rapid and continuous power control capabilities [17]. Furthermore, to accommodate the uncertainties associated with high renewable energy integration, dynamic reconfiguration and voltage control strategies based on model predictive control and robust optimization have gained prominence. These approaches are particularly effective in mitigating violations over shorter operational time scales [18]-[20].

When operational controls prove insufficient to eliminate overloads, infrastructure reinforcement at the planning level becomes imperative. This challenge is usually formulated as an optimization problem aimed at minimizing total investment and operational costs [21]. Some common upgrade measures, such as reconductoring or constructing new lines, provide a direct increase in physical capacity but often entail prohibitive capital costs [22]. To enhance economic efficiency, recent literature has extensively investigated hybrid frameworks that integrate battery energy storage systems (BESS) into the planning process [23]. By co-optimizing BESS allocation with line upgrades, utilities can defer costly infrastructure investments, leveraging the load-shifting capabilities of storage to mitigate thermal overloads during peak periods [24]. By integrating a refined battery degradation model, researchers evaluated the trade-off between initial capital investment and long-term operational and lifecycle costs. The findings demonstrate that planning strategies explicitly accounting for state of health dynamics significantly enhance the overall economic efficiency of the project [25].

Coordinated charging of EVs is emerging as a compelling non-wire alternative for mitigating distribution grid congestion [26]. Literature indicates that optimized coordination strategies can effectively shift large-scale EV loads away from peak periods [27]. Furthermore, vehicle-to-grid (V2G) capabilities enable EVs to provide active grid support, thereby resolving thermal overloads without necessitating costly upgrades to transformer and line capacities [28]. However, a significant challenge of the high uncertainty of user behavior remains. To address this issue, recent studies have increasingly adopted robust optimization and data-driven methodologies to model these stochastic factors, ensuring strategic efficacy even under worst-case scenarios [29].

In summary, addressing thermal overloads and voltage violations in distribution networks necessitates a balanced evaluation of diverse mitigation strategies. While EV charging coordination offers a cost-effective and flexible non-wire alternative, its efficacy is limited by the stochastic nature of user participation. However, the installation of BESS provides rapid response capabilities but is constrained by high capital costs and degradation concerns, necessitating rigorous lifecycle analysis. The network reinforcement remains the definitive solution for physical capacity expansion yet is characterized by significant implementation delays and capital expenditure. Therefore, emerging research is shifting towards a joint planning and optimization framework. This framework aims to achieve overall technical and economic optimality by utilizing flexible resources to manage peak demand while maintaining the structural integrity of the power grid.

To address critical gaps in prior research, this study provides an in-depth assessment of line loading levels, thermal overloads, and voltage deviations under various EV charging scenarios. The main contributions of this paper are summarized as follows:

- This study rigorously investigates system violations under varying EV penetration rates from 0% to 100% in 20% increments and different Level-2 charging capacities with 5, 10, and 15 kW. Furthermore, it uniquely quantifies the shift in critical EV hosting capacity thresholds before and after BESS integration.
- The paper formulates the Violation Detection and Quantification (VDQ) model to systematically pinpoint

network breaches, and the Violation Mitigation-Based Planning (VMBP) model to optimize battery energy storage system allocations.

- To overcome the computational intractability of directly combining cable upgrades and BESS installations, a novel three-stage DDCP framework is proposed. It strategically decouples the joint optimization by: (1) diagnosing bottleneck lines that render standalone BESS solutions infeasible, (2) executing targeted physical upgrades exclusively on these lines, and (3) optimizing BESS deployment within the newly enhanced network.
- The techno-economic superiority of the proposed DDCP framework is comprehensively validated by comparing it with three alternative mitigation approaches: the VDQ-driven cable upgrade (VCU) model, the VMBP model and system-wide voltage uprating.

## II. The Proposed DDCP Framework

This section proposes a diagnosis-driven co-planning (DDCP) framework to efficiently solve the violation issues by combining the advantages of cable upgrades and BESS deployment. The framework develops two hourly-resolution distribution-flow models: the violation detection and quantification (VDQ) model to systematically identify system violations, and the violation-mitigated BESS planning (VMBP) model to mitigate line thermal capacity and voltage violations. Initially, the VDQ model assesses system constraints under different reference voltage levels and EV charger capacities, explicitly capturing violations induced by fluctuating EV penetration. Subsequently, the VMBP model determines the optimal sizing and siting of BESS installations to resolve the identified network violations.

### *A. Violation Detection and Quantification Model*

The VDQ model is designed in this paper to detect and quantify the line ampacity and voltage violations, using a distribution-flow model based on the assumption of a radial network. The objective is to minimize the active power losses as shown in (1):

$$objective = min\left(\sum_{(i,j)\in L}\sum_{t\in T} r_{ij}\cdot l_{ij}^{t}\right) \quad (1)$$

where $L$ is the set of branches and $r_{ij}$ represents the resistance of the line connecting nodes $i$ and $j$.

The nodal active and reactive power balance equations at non-substation buses and substation bus are expressed as (2) and (3), respectively.

$$\begin{cases}\sum_{k\in U(i)}(P_{ki}^{t}-r_{ki}\cdot l_{ki}^{t})=\sum_{j\in D(i)}P_{ij}^{t}+P_{load,i}^{t}\\ \sum_{k\in U(i)}(Q_{ki}^{t}-x_{ki}\cdot l_{ki}^{t})=\sum_{j\in D(i)}Q_{ij}^{t}+Q_{load,i}^{t}\end{cases} \quad (2)$$
$$,\forall i\in N,\forall t\in T$$

$$\begin{cases}\sum_{k\in U(s)}(P_{ks}^{t}-r_{ks}\cdot l_{ks}^{t})=-P_{s}^{t}+\sum_{j\in D(s)}P_{sj}^{t}+P_{load,s}^{t}\\ \sum_{k\in U(s)}(Q_{ks}^{t}-x_{ks}\cdot l_{ks}^{t})=-Q_{s}^{t}+\sum_{j\in D(s)}Q_{sj}^{t}+Q_{load,s}^{t}\end{cases} \quad (3)$$
$$,\forall s\in G,\forall t\in T$$

where for each node $i\in N$, $D(i)$ and $U(i)$ represent the sets of its downstream and upstream nodes, respectively.

The voltage drops across a line can be calculated in (4), and the voltage constraint of the substation bus is defined by (5):

$$v_{i}^{t}-v_{j}^{t}=2\left(r_{ij}P_{ij}^{t}+x_{ij}Q_{ij}^{t}\right)-\left(r_{ij}^{2}+x_{ij}^{2}\right)l_{ij}^{t}$$
$$,\forall(i,j)\in L,\forall t\in T \quad (4)$$

$$v_{s}^{t}=(V_{s}^{t})^{2},\forall s\in G,\forall t\in T \quad (5)$$

The second-order cone constraint between line current and power can be defined as:

$$v_{i}^{t}\cdot l_{ij}^{t}\geq(P_{ij}^{t})^{2}+(Q_{ij}^{t})^{2},\forall(i,j)\in L,\forall t\in T \quad (6)$$

The relaxed constraints of voltage and current are shown in (7)-(8), and $I_{ij}^{max}$ is the per-unit maximum current ampacity between node $i$ and $j$.

$$v_{i}^{t}\leq v^{max},\forall i\in N,\forall t\in T \quad (7)$$

$$0\leq l_{ij}^{t},\forall(i,j)\in L,\forall t\in T \quad (8)$$

To assess the loading levels and potential violations in the system, the line loading is computed as the ratio of the actual current $I_{ij}^{t}$ to the rated current as formulated in (9). If the line loading exceeds 100%, a violation occurs, indicating that the line is carrying more current than the rated ampacity.

$$\text{Line Loading}=\frac{I_{ij}^{t}}{I_{ij}^{max}}\times 100\%,\forall(i,j)\in L,\forall t\in T \quad (9)$$

The severity of the violation is quantified by calculating the violation percentage, which is determined by (10). A higher violation percentage indicates a greater deviation from the rated ampacity. This calculation helps identify areas where the grid may be overburdened and requires corrective action to prevent potential failures or damage. If the voltage value exceeds the acceptable range, it constitutes a voltage violation.

$$\text{Violation }=\left[\frac{I_{ij}^{t}}{I_{ij}^{max}}-1\right]_{+}*100\% \quad (10)$$

where the definition of positive part function is $[x]_{+}\triangleq\max(x,0)$.

In summary, the VDQ model aims to evaluate the grid impact by optimizing the objective function defined in (1), subject to the power flow and operational constraints specified in equations (2)-(10).

### *B. VDQ-Driven Cable Upgrade Model*

Violation issues identified by the VDQ model can be resolved through cable upgrades in the VDQ-driven cable upgrade (VCU) model. This approach is effective because upgraded cables simultaneously enhance the current-carrying capacity and decrease the line impedance, thereby mitigating voltage drops. Built upon the VDQ model, its original parameters impedance $r_{ij}$ , $x_{ij}$ and ampacity $I_{ij}^{max}$ of the overloaded lines are converted into decision variables $r_{c}$, $x_{c}$ and $I_{c}^{max}$. A binary variable $z_{ij,c}$ is introduced to denote the selection of the $c$-th cable type for line between node $i$ and $j$. Each line within the cable upgrade candidate set may contain at most one cable upgrade option, defined as follows:

$$\sum_{c\in C_{ij}}z_{ij,c}\leq 1,\forall(i,j)\in\Omega_{cand} \quad (11)$$

where $\Omega_{cand}$ is the set of upgrade candidate lines consisting of the overloaded lines, and $C_{ij}$ is the set of upgrade cable types that can be selected between node $i$ and $j$.

The line capacity constraints change dynamically according to the upgraded cable in (12) where $\sigma_{ij}^{t}$ is the slack variable.

$$0 \le l_{ij}^t \le (I_{ij}^{max})^2 \cdot \left(1-\sum_{c\in C_{ij}} z_{ij,c}\right) + \sum_{c\in C_{ij}} (I_c^{max})^2 \cdot z_{ij,c} + \sigma_{ij}^t \quad ,\forall (i,j)\in\Omega_{cand}, \forall t \in T \tag{12}$$

To maintain the convexity of the optimization model and avoid nonlinear bilinear terms, the Big M method will be used to reconstruct the original voltage drop equation (4) and network loss constraint (2) as follows:

$$-M(1-z_{ij,c}) \le v_i^t - v_j^t - 2\left(r_c P_{ij}^t + x_c Q_{ij}^t\right) + (r_c^2 + x_c^2) l_{ij}^t \le M(1-z_{ij,c}), \forall (i,j)\in\Omega_{cand}, \forall c\in C_{ij}, \forall t\in T \tag{13}$$

$$-M(1-z_{ij,c}) \le P_{ij}^{loss,t} - r_c l_{ij}^t \le M(1-z_{ij,c}) \quad ,\forall (i,j)\in\Omega_{cand}, \forall c\in C_{ij}, \forall t\in T \tag{14}$$

$$-M(1-z_{ij,c}) \le Q_{ij}^{loss,t} - x_c l_{ij}^t \le M(1-z_{ij,c}) \quad ,\forall (i,j)\in\Omega_{cand}, \forall c\in C_{ij}, \forall t\in T \tag{15}$$

$$\begin{cases} \sum_{k\in U(i)} (P_{ki}^t - P_{ki}^{loss,t}) = \sum_{j\in D(i)} P_{ij}^t + P_{load,i}^t \\ \sum_{k\in U(i)} (Q_{ki}^t - Q_{ki}^{loss,t}) = \sum_{j\in D(i)} Q_{ij}^t + Q_{load,i}^t \end{cases} ,\forall i\in N, \forall t\in T \tag{16}$$

where $M$ is a large positive number and directly constructing the voltage drop and network loss equations necessitates the multiplication of discrete impedance variables with continuous flow variables or continuous line capacity variable, yielding non-convex bilinear terms. To avoid this intractability and preserve model convexity, this paper introduces continuous auxiliary variables $P_{ij}^{loss,t}$ and $Q_{ij}^{loss,t}$ for the branch losses in constraints (14)-(16). Combined with the Big-M method, the nonlinear multiplications are exactly linearized into a set of mixed-integer linear constraints. However, the selection of the value for $M$ has a significant impact on both the algorithmic speed and numerical precision of the solver. If $M$ is chosen to be excessively large, it may severely undermine the effectiveness of the linear programming relaxation, thereby causing the solving time to expand exponentially, consequently reducing computational efficiency. If $M$ is chosen to be too small, there is a risk of excluding the true optimal solution. Therefore, the value of $M$ is strictly derived based on the physical upper bounds within the specific linearization context. Meanwhile, the voltage constraints for all nodes, as well as the strict constraints on line currents for lines not included in the set of upgrade candidate lines $\Omega_{cand}$, are defined by (17)-(18):

$$v^{min} \le v_i^t \le v^{max}, \forall i\in N, \forall t\in T \tag{17}$$

$$0 \le l_{ij}^t \le (I_{ij}^{max})^2, \forall (i,j)\in L\backslash\Omega_{cand}, \forall t\in T \tag{18}$$

And the objective will be updated to minimize the cost of upgrading cables and the penalty for slack variables in (19).

$$objective = \min\left(\begin{array}{c} \sum_{(i,j)\in\Omega_{cand}} \sum_{c\in C_{ij}} Cost_c \cdot z_{ij,c} \\ + \sum_{(i,j)\in\Omega_{cand}} \sum_{t\in T} \rho \cdot \sigma_{ij}^t \end{array}\right) \tag{19}$$

where $Cost_c$ represents the cost required to upgrade with the $c$-th cable type, and $\rho$ is a large positive penalty number.

To conclude, the VCU benchmark model seeks to minimize the traditional grid-reinforcement cost through the objective function (19), governed by the constraints outlined in (3), (5)-(6), and (11)-(18).

### *C. Violation Mitigation-Based Planning Model*

Unlike the VCU model to address identified violations, the VMBP model is developed as a planning model for co-optimizing BESS installation capacity and location to solve violation issues. The objective function of the VMBP model is to minimize the BESS installation cost.

$$objective = min\left(\sum_{b\in B} c^{cap} \cdot E_b^{cap}\right) \tag{20}$$

where $c^{cap}$ represents the capacity cost of the BESS.

The binary decision variable $z_b$ represents the installation status of a BESS at bus $b$, $y_{b,t}^{charge}$ and $y_{b,t}^{discharge}$ indicate the active power charging and discharging states and $y_{b,t}^{inj}$ and $y_{b,t}^{abs}$ denote the status of reactive power injection and absorption. The upper and lower limits of the BESS capacity are defined in constraint (22), and constraints (23)-(24) prevent BESS from charging and discharging, or injecting and absorbing reactive power simultaneously.

$$z_b, y_{b,t}^{inj}, y_{b,t}^{abs}, y_{b,t}^{charge}, y_{b,t}^{discharge} \in \{0,1\} \tag{21}$$

$$z_b \cdot E_{min}^{cap} \le E_b^{cap} \le z_b \cdot E_{max}^{cap}, \forall b\in B \tag{22}$$

$$y_{b,t}^{inj} + y_{b,t}^{abs} \le 1, \forall b\in B, \forall t\in T \tag{23}$$

$$y_{b,t}^{charge} + y_{b,t}^{discharge} \le 1, \forall b\in B, \forall t\in T \tag{24}$$

$E_{b,t}^{bess}$ is constrained by the operational upper and lower bounds in (25). The SOC is modeled as a temporally coupled variable in (26), describing its evolution based on the previous state and the current net charging or discharging power. To ensure sustainable cyclical operation, the model constrains a boundary condition requiring the SOC at the initial time step $t_0$ to equal that at the last time step $t_{end}$ as shown in (27).

$$SOC^{min} \cdot E_b^{cap} \le E_{b,t}^{bess} \le SOC^{max} \cdot E_b^{cap}, \forall b\in B, \forall t\in T \tag{25}$$

$$E_{b,t}^{bess} = E_{b,t-1}^{bess} + \left(P_{b,t}^{BESS_{ch}} \eta_{charge} - \frac{P_{b,t}^{BESS_{dis}}}{\eta_{discharge}}\right) ,\forall b\in B, \forall t\in T\backslash\{t_0\} \tag{26}$$

$$E_{b,t_0}^{bess} = E_{b,t_{end}}^{bess}, \forall b\in B \tag{27}$$

To circumvent the computational complexity of quadratic constraints, a Big-M formulation is adopted to linearize the power limits in (28)-(33). In these expressions, $M$ represents a sufficiently large positive constant, chosen to ensure that the constraints remain non-binding when the corresponding binary variables are inactive.

$$0 \le P_{b,t}^{BESS_ch} \le C_{rate}^{ch} \cdot E_b^{cap}, \forall b\in B, \forall t\in T \tag{28}$$

$$0 \le P_{b,t}^{BESS_ch} \le M \cdot y_{b,t}^{charge}, \forall b\in B, \forall t\in T \tag{29}$$

$$0 \le P_{b,t}^{BESS_dis} \le C_{rate}^{dis} \cdot E_b^{cap}, \forall b\in B, \forall t\in T \tag{30}$$

$$0 \le P_{b,t}^{BESS_dis} \le M \cdot y_{b,t}^{discharge}, \forall b\in B, \forall t\in T \tag{31}$$

$$0 \le Q_{b,t}^{BESS_inj} \le M \cdot y_{b,t}^{inj}, \forall b\in B, \forall t\in T \tag{32}$$

$$0 \le Q_{b,t}^{BESS_abs} \le M \cdot y_{b,t}^{abs}, \forall b\in B, \forall t\in T \tag{33}$$

The inverter capacity of BESS is defined in (34). The relationship between active power, reactive power, and the apparent power of the inverter is given by (35).

$$S_{inv,b} = K_{inv} \cdot C_{rate}^{dis} \cdot E_b^{cap}, \forall b\in B \tag{34}$$

$$\left(P_{b,t}^{BESS_{ch}} + P_{b,t}^{BESS_{dis}}\right)^2 + \left(Q_{b,t}^{BESS_{abs}} + Q_{b,t}^{BESS_{inj}}\right)^2 \le S_{inv,b}^2 \quad ,\forall b\in B, \forall t\in T \tag{35}$$

Nodal power balance equations for non-substation nodes considering BESS are defined in (36) and (37).

$$\sum_{k\in U(i)} (P_{ki}^t - r_{ki} \cdot l_{ki}^t) = \sum_{j\in D(i)} P_{ij}^t + P_{load,i}^t + \sum_{b\in B_i} \left(P_{b,t}^{BESS_ch} - P_{b,t}^{BESS_dis}\right), \forall i\in N, \forall t\in T \tag{36}$$

$$\sum_{k\in U(i)} (Q_{ki}^t - x_{ki}\cdot l_{ki}^t) = \sum_{j\in D(i)} Q_{ij}^t + Q_{load,i}^t$$
$$+ \sum_{b\in B_i} (Q_{b,t}^{BESS_abs} - Q_{b,t}^{BESS_inj}), \forall i \in N, \forall t \in T \qquad (37)$$

In summary, the VMBP model focuses on minimizing the BESS installation cost as expressed in the objective function (20), strictly adhering to the grid and storage operational constraints defined in (3)-(6), (17)-(18), and (21)-(37).

### *D. Diagnosis-Driven Co-planning (DDCP) Framework*

Although cable upgrading is a proven method for mitigating thermal violations, it is very expensive and time-consuming for utility implementation. Similarly, the VMBP model cannot fully resolve profound cable overload issues. Moreover, the direct joint optimization of large-scale cable upgrades and BESS deployment introduces extreme computational intractability driven by the curse of dimensionality. To address these intractability challenges, this paper proposes the DDCP framework which is a three-stage collaborative optimization strategy tailored to decompose and efficiently solve the complex joint planning problem. The DDCP framework process is shown in Fig. 1. The interface information flows sequentially across the framework: Stage I identifies critical bottleneck lines $\Omega_{\text{diag}}$, Stage II upgrades them to generate updated network parameters as $I_{ij,max}^{upg}$, and Stage III utilizes this enhanced topology to finalize BESS placement and sizing.

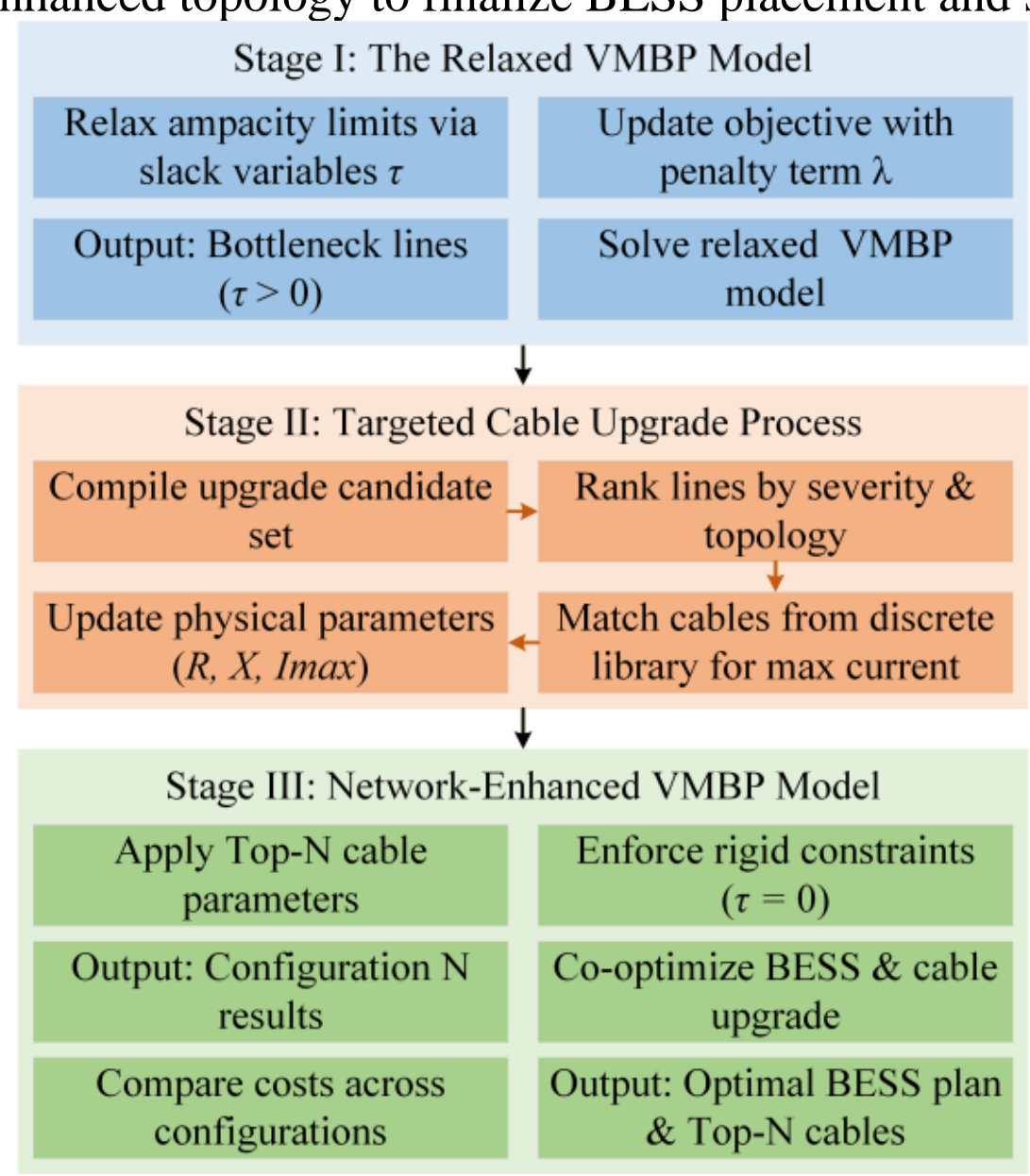

Fig. 1. Flowchart of the DDCP framework.

#### *Stage I: The Relaxed VMBP Model*

When the line is severely overloaded, simply installing a BESS not only fails to solve the overload problem but also makes it difficult to satisfy the constraint (27) in the VMBP model. To make the optimization model feasible and diagnose the bottleneck lines with severe current overload, it is necessary to relax the upper limit of the line current carrying ampacity. Therefore, a slack variable $\tau_{ij}^t$ is introduced into the current carrying ampacity constraint, converting constraint (8) into (38), and the corresponding penalty term and its penalty coefficient λ are included in the updated objective function (39). This form enables the solver to identify the specific bottleneck lines that cause the system to be infeasible.

$$0 \le l_{ij}^t \le (I_{ij}^{max})^2 + \tau_{ij}^t, \forall t \in T, \forall (i,j) \in L \qquad (38)$$

$$objective = min\left(\sum_{b\in B} c^{cap}\cdot E_b^{cap}\right) + \lambda \sum_{(i,j)\in L, t\in T} \tau_{ij}^t \qquad (39)$$

In contrast to physical cable upgrades that resolve violation issues by altering line ampacity and impedance, Stage I is strictly diagnostic based on the relaxed VMBP model. The primary objective is to identify the critical bottleneck lines that render the BESS planning model infeasible, thereby restoring the solvability of the optimization problem. The slack variable $\tau_{ij}^t$ quantifies the residual current exceeding the limit that still exists on the line between node $i$ and $j$ at time $t$ after the optimal configuration of BESS. Based on these diagnostic results, a candidate upgrade set $\Omega_{\text{diag}}$ is defined to aggregate all lines exhibiting such irreducible violations at any given time:

$$\Omega_{\text{diag}} = \{(i,j) \in L \mid \sum_{t\in T} \tau_{ij}^t > 0\} \qquad (40)$$

#### *Stage II: Targeted Cable Upgrade Process*

Utilizing the diagnostic results from Stage I, the identified bottleneck lines are compiled into an upgrade set and prioritized in descending order based on their violation severity and network connectivity. A discrete library of standard commercial cables is introduced as candidate new cables. For each targeted line, appropriate replacement cable is selected to accommodate the maximum current flow observed during Stage I of constraint relaxation. The current carrying ampacity constraint for these targeted lines is updated as:

$$0 \le l_{ij}^t \le (I_{ij,max}^{upg})^2, \forall (i,j) \in \Omega_{\text{diag}}, \forall t \in T \qquad (41)$$

where $I_{ij,max}^{upg}$ represents the new current limit of the upgraded lines. Besides, upgrading cables not only increases the thermal ampacity but also lowers the resistance and reactance parameters to mathematically reflect the physical cross-sectional area of the selected new conductor.

#### *Stage III: Network-Enhanced VMBP Model*

In Stage III, the network-enhanced VMBP model incorporates the Top-N prioritized cable upgrades from Stage II by modifying the corresponding branch impedances and ampacity limits. With the network physical parameters updated, the slack variables from Stage I are removed to enforce rigid system constraints. The optimal deployment location and capacity of the BESS are then optimized. By conducting a comparative analysis across various N configurations, the proposed framework successfully determines the optimal BESS locations, capacities, and operational dispatch.

In summary, the objectives and constraints of the five optimization models are shown in Table I. The functions and coordination relationships of the five optimization models are summarized as follows:

- VDQ Model: The base power flow model used to assess EV charging impacts and identify voltage and current violations.
- VCU Model: A conventional benchmark that mitigates the identified violations from VDQ model through cable upgrades only.

- VMBP Model: A BESS-only benchmark that attempts to eliminate the violations from the VDQ model through optimized storage placement and sizing.
- Relaxed VMBP Model (Stage I of DDCP): A diagnostic formulation used when BESS alone cannot fully restore feasibility. It relaxes line current limits to reveal the critical bottleneck lines.
- Network-Enhanced VMBP Model (Stage III of DDCP): The final co-planning model, where the identified bottleneck lines are upgraded and incorporated into the network to optimize BESS placement and sizing while eliminating all remaining violations.

TABLE I. COMPARISON OF OBJECTIVES AND CONSTRAINTS OF DIFFERENT MODELS

| Model | Objective | Constraints |
|---|---|---|
| VDQ model | (1) | (2)-(10) |
| VCU model | (19) | (3), (5)-(6), (11)-(18) |
| VMBP model | (20) | (3)-(6), (17)-(18), (21)-(37) |
| Relaxed VMBP model | (39) | (3)-(6), (17), (21)-(38) |
| Network-Enhanced VMBP model | (20) | (3)-(6), (17), (21)-(37), (40)-(41) |

## III. EV LOAD IMPACT RESULTS

This section utilizes the VDQ model to statistically analyze and quantify network limit violations, specifically line thermal overloads and nodal voltage deviations. The comprehensive assessment is conducted across five base voltage levels (4.16 kV, 6.9 kV, 13.8 kV, 23.9 kV, and 34.5 kV [30]) and three EV charger capacities (5 kW, 10 kW, and 15 kW), evaluated under progressive EV penetration levels ranging from 0% to 100% in 20% increments.

### A. Test System Design and Description

The test system employed in this study, as detailed in our preliminary work [1], utilizes a real-world 240-bus distribution system located in the Midwest U.S., which serves 1,120 customers via three primary feeders and 23 miles of conductors. The test framework, illustrated in Fig. 2, integrates one year of smart meter data from 2017 [31] to identify worst-case loading conditions, specifically the peak demand occurring at 13:00 on July 12. EV charging loads are probabilistically distributed across the network using the Mersenne Twister algorithm, ensuring that charger allocation is proportional to the customer density at each bus [32]. This study strictly assesses the worst-case impacts of EV charging loads. During the static grid impact analysis, maximum rated EV loads are superimposed directly onto the base peak hour at 13:00 on July 12. For the optimization of mitigation solutions, this maximum EV load is applied as a fixed continuous 24-hour stress-test envelope over the baseline load on July 12. Such worst-case consideration ensures that the resulting infrastructure and BESS capacities are robust against extreme and sustained grid congestion. The impact of varying EV adoption rates and Level-2 charger capacities is then evaluated across multiple distribution voltage levels using the VDQ model to obtain the branch power flows and bus voltages. To facilitate modeling, the 240-bus system is approximated as a modified balanced three-phase network by converting single-phase line configurations into three-phase equivalents. This approximation assumes that the self-impedance of all phases is identical and that the mutual inductance is symmetrically distributed. Furthermore, it assumes that both the base load and the EV charging load are distributed completely evenly across each phase, thereby neglecting the neutral current and the voltage drop resulting from it. These assumptions are adopted primarily to maintain the computational tractability of the large-scale mixed-integer linear programming (MILP) formulation. While providing an efficient and robust planning baseline, it is acknowledged that such approximations may slightly underestimate the localized voltage drops on the most heavily loaded phases when applied to severely unbalanced real-world feeders.

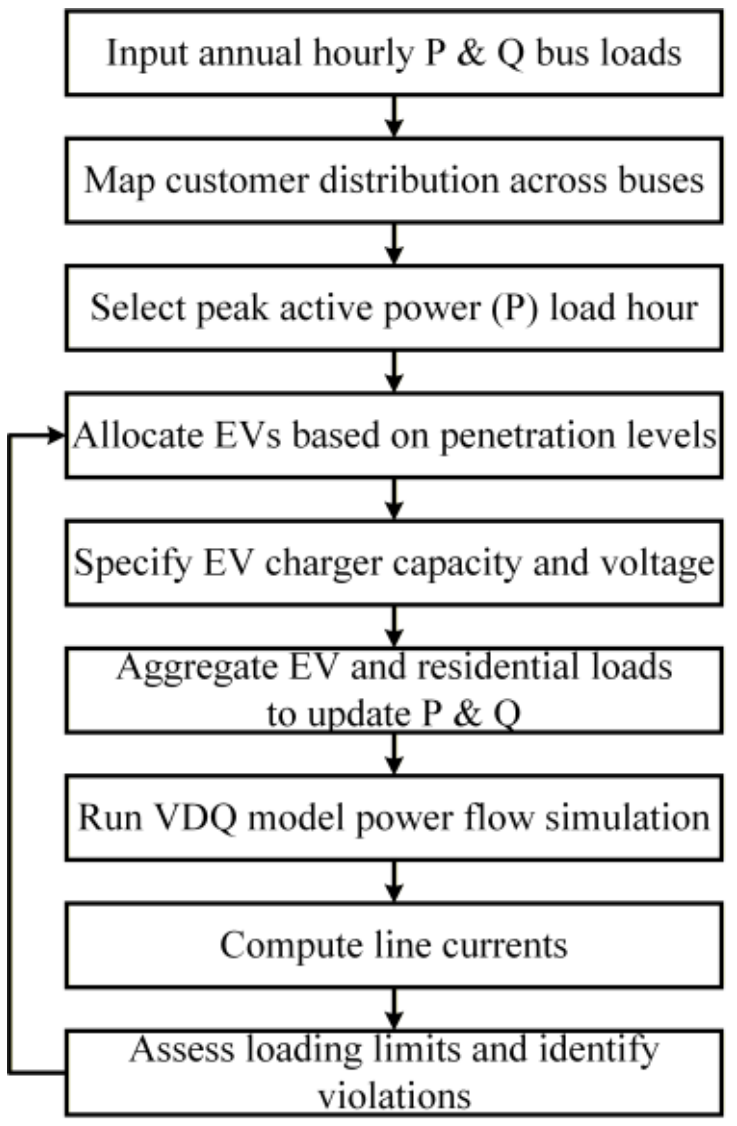

Fig. 2. Flowchart of the EV load integration and grid impact assessment procedure.

### B. Visualization of EV-Induced Grid Congestions

After calculating line currents across various voltage levels and EV penetration rates, we highlighted the line violations in Fig. 3 for a system voltage of 6.9 kV and an EV charger capacity of 10 kW [15]. A cumulative color-coding approach is employed in the legend to represent these violations at different EV adoption rates.

For instance, blue marks lines that experience violations at a 20% EV rate, while yellow indicates additional lines that become overloaded at 40% penetration, also encompassing those previously overloaded at 20%. This pattern continues, with each subsequent color representing an increasing penetration level and the cumulative effect on the network. The illustration reveals that lines closest to the substation are most vulnerable to overloading, as they carry the highest cumulative power flow from downstream buses. As EV penetration increases, the stress extends progressively further along the network, with each subsequent line having a higher chance of being overloaded. This cascading effect emphasizes the strain on distribution systems as EV adoption grows.

### C. Statistical Overview of Baseline Violations

This section presents a statistical comparison of the current and voltage violations induced by the integration of 10 kW EV chargers across different distribution network voltage levels. The thermal heatmap in Fig. 4 shows the maximum loading levels across various voltage levels for different EV adoption rates with 10-kW chargers. Even at a 0% EV adoption rate, line overloads persist when the base voltage is set to 4.16 kV. When the EV adoption rate exceeds 20%, the model will collapse. At the 6.9 kV level, the maximum loading percentage increases sharply from 59.91% with no EVs to 306.26% at full adoption, signaling a high overload risk. At higher voltages, such as 13.8 kV and 34.5 kV, the system's resilience improves. The maximum loading at 13.8 kV is 126.59% and at 34.5 kV is 48.83%. This highlights the growing strain at higher EV adoption rates and the improved capacity at higher voltage levels.

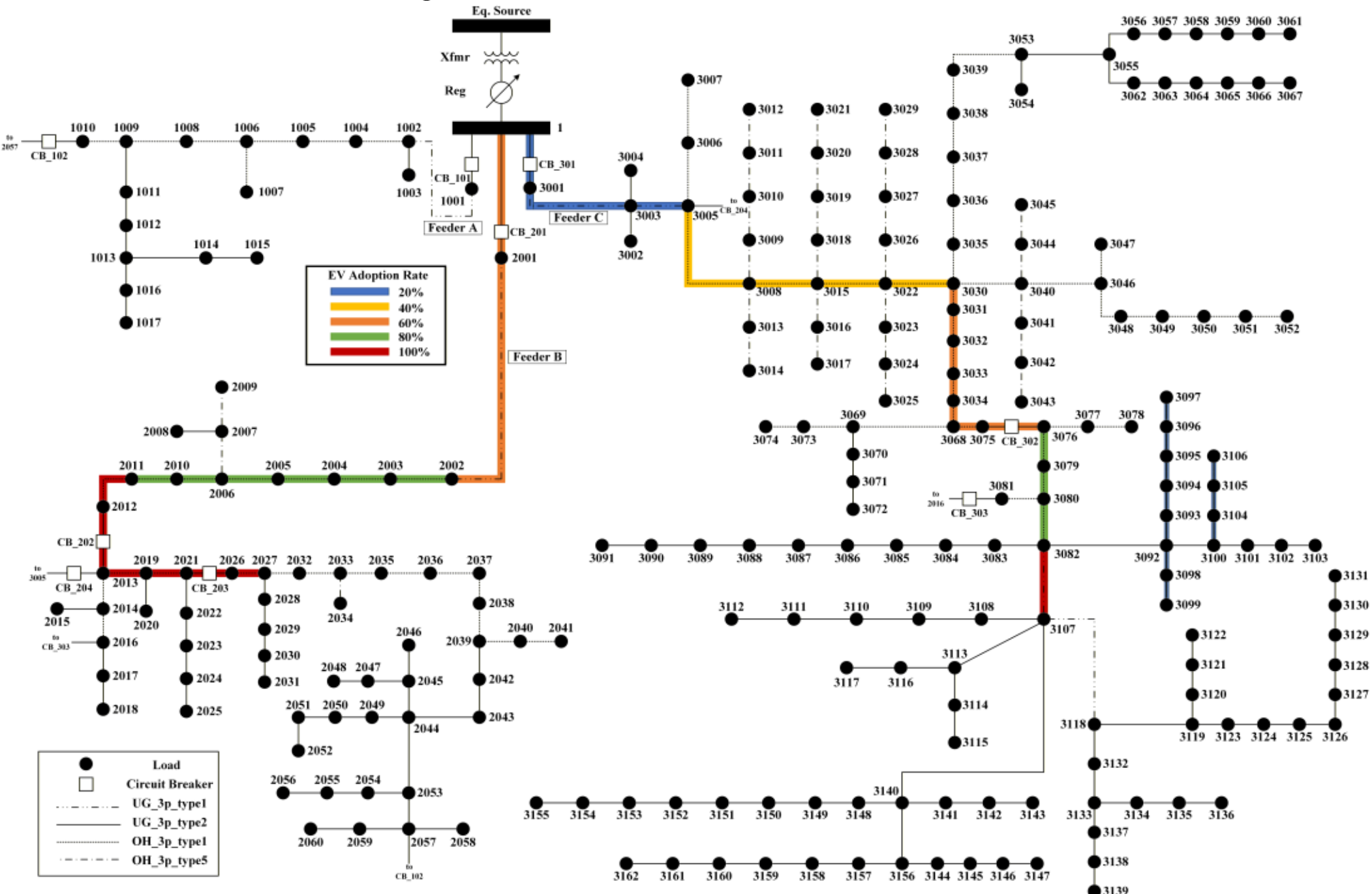

Fig. 3. Line Violations for different EV penetration rates given distribution voltage level of 6.9 kV, assuming the charging power is 10 kW.

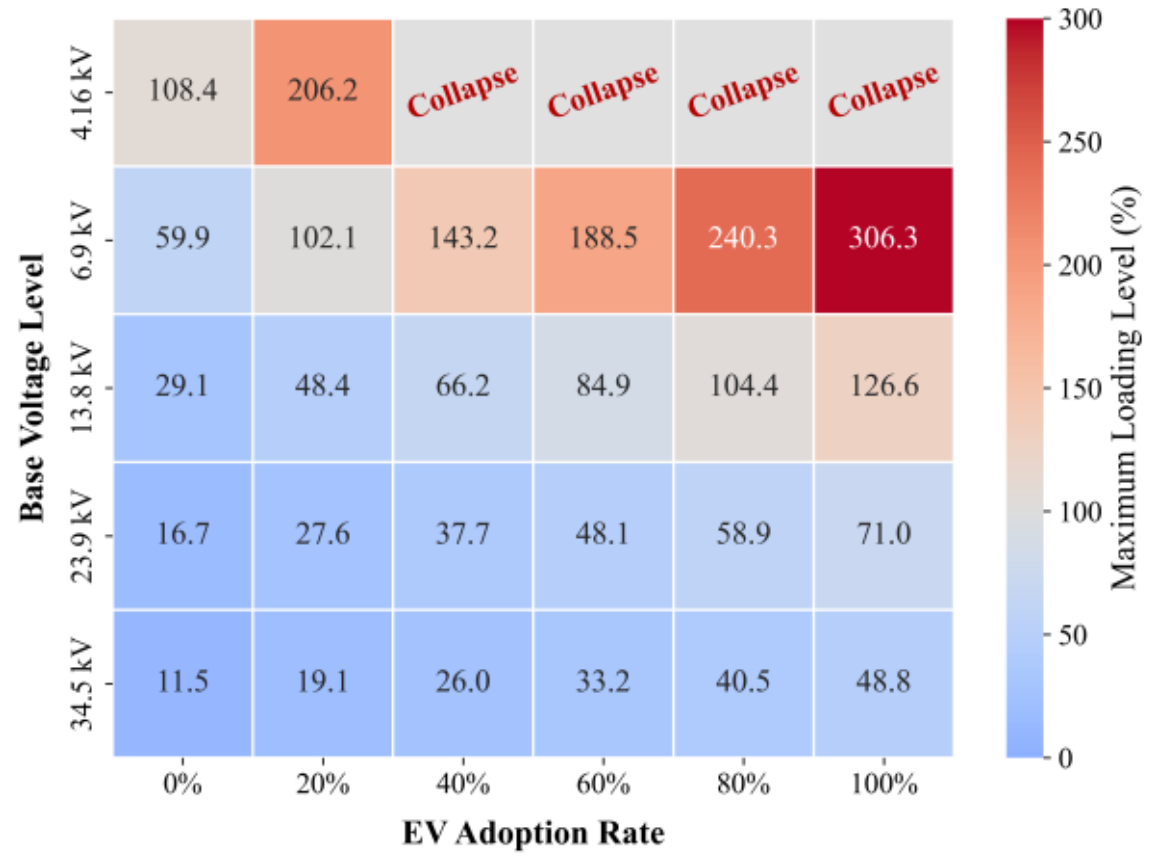

Fig. 4. Impact of EV integration on maximum system loading level.

TABLE II. LINE VIOLATIONS STATISTICS FOR EV CHARGER P = 10 KW

| Voltage Level | Violation Metric | EV Adoption Rate | | | | | |
|---|---|---|---|---|---|---|---|
| | | 0% | 20% | 40% | 60% | 80% | 100% |
| 4.16 kV | Count | 4 | 16 | Model collapse | | | |
| | Min % | 0.37 | 1.07 | | | | |
| | Max % | 8.43 | 106.2 | | | | |
| | Avg % | 5.10 | 49.56 | | | | |
| 6.9 kV | Count | 0 | 2 | 7 | 14 | 20 | 30 |
| | Min % | | 1.08 | 12.19 | 1.78 | 2.58 | 0.5 |
| | Max % | | 2.12 | 43.22 | 88.46 | 140.3 | 206.3 |
| | Avg % | | 1.60 | 31.15 | 38.52 | 58.62 | 70.70 |
| 13.8 kV | Count | 0 | 0 | 0 | 0 | 3 | 6 |
| | Min % | | | | | 0.11 | 4.92 |
| | Max % | | | | | 4.41 | 26.59 |
| | Avg % | | | | | 2.72 | 17.91 |
| 23.9 kV | Count | 0 | | | | | |
| 34.5 kV | Count | 0 | | | | | |

Table II demonstrates the line violations where any loading exceeding 100% is considered a violation. At 4.16 kV, violations increase from 8.43% at 0% adoption to 106.2% at 20% adoption. At 6.9 kV, violations worsen with higher EV adoption, reaching 206.26% at full adoption, with violations rising from 2 to 30 instances. The 13.8 kV level shows fewer violations, peaking at 26.59% at full adoption, while 23.9 kV and 34.5 kV show no violations, indicating better capacity for handling EV loads. This reinforces the trend that higher voltage levels offer stronger infrastructure for EV integration.

Table III summarizes the statistical results of nodal per-unit voltages across various base voltage levels, assuming a constant EV charger power of 10 kW. A voltage violation is defined as any node voltage falling below 0.95 p.u. At the 4.16 kV level, the system is highly sensitive and the number of violations rises from 158 at 0% EV adoption to 197 at 20% adoption, with the minimum voltage dropping to 0.73 p.u. Similarly, at 6.9 kV, violations increase from 86 at 20% adoption to 195 at 100% adoption, with a minimum voltage of 0.76 p.u. In contrast, no voltage violations were observed at higher base voltage levels.

TABLE III. NODAL VOLTAGE STATISTICS FOR EV CHARGER P = 10 KW

| Voltage Level | Voltage Metric | EV Adoption Rate | | | | | |
|---|---|---|---|---|---|---|---|
| | | 0% | 20% | 40% | 60% | 80% | 100% |
| 4.16 kV | Count | 158 | 197 | Model collapse | | | |
| | Min (p.u.) | 0.87 | 0.73 | | | | |
| | Avg (p.u.) | 0.92 | 0.84 | | | | |
| 6.9 kV | Count | 0 | 86 | 155 | 175 | 184 | 195 |
| | Min (p.u.) | 0.97 | 0.93 | 0.90 | 0.86 | 0.82 | 0.76 |
| | Avg (p.u.) | 0.99 | 0.97 | 0.94 | 0.92 | 0.89 | 0.86 |
| 13.8 kV | Count | 0 | | | | | |
| 23.9 kV | Count | 0 | | | | | |
| 34.5 kV | Count | 0 | | | | | |

### *D. Analysis Under Diverse Network Configurations*

This subsection provides a detailed investigation of the operational violations induced by different EV charger capacities and penetration rates at a base voltage of 6.9 kV. In addition, scatter plots are provided to visualize the nodal voltage distribution and branch loading percentages across the 4.16 kV, 6.9 kV, and 13.8 kV systems under all considered EV scenarios.

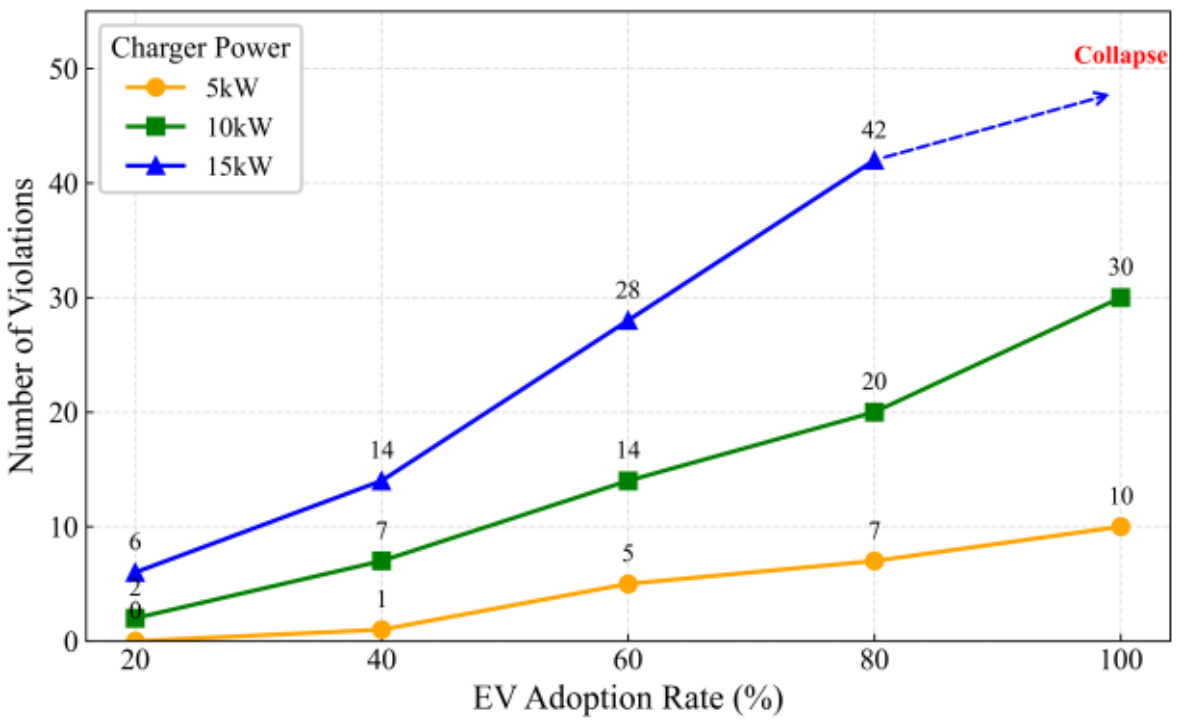

Fig. 5. Line violation count for varying power capacities at 6.9 kV.

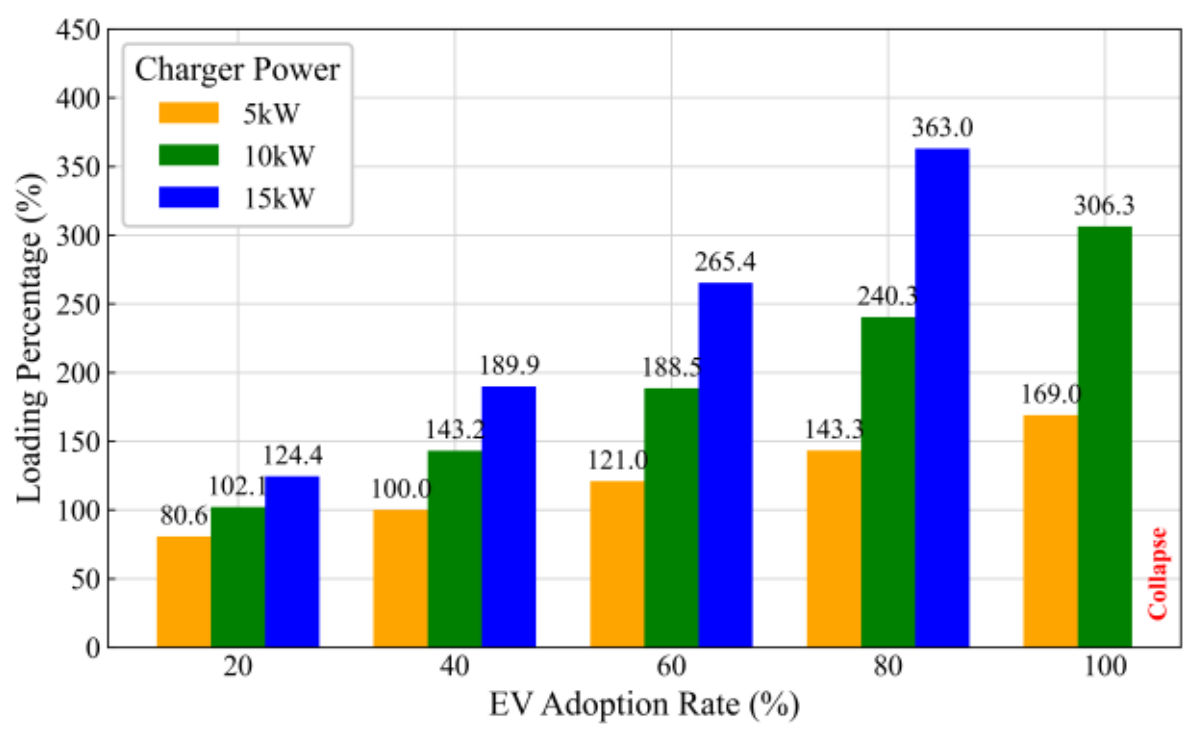

Fig. 6. Max loading percentage for varying power capacities at 6.9 kV.

A sensitivity analysis of EV charger power levels (5 kW, 10 kW, and 15 kW) shows that higher-power chargers lead to more line violations as adoption increases. Fig. 5 shows 15 kW chargers cause 14 violations at 40% adoption, while 5 kW chargers result in 10 violations at 100% adoption. Fig. 6 reveals that higher charger power and adoption increase loading levels, with maximum loading reaching 363.0% at 15 kW and 80% adoption, compared to 306.3% at 10 kW and 100% adoption.

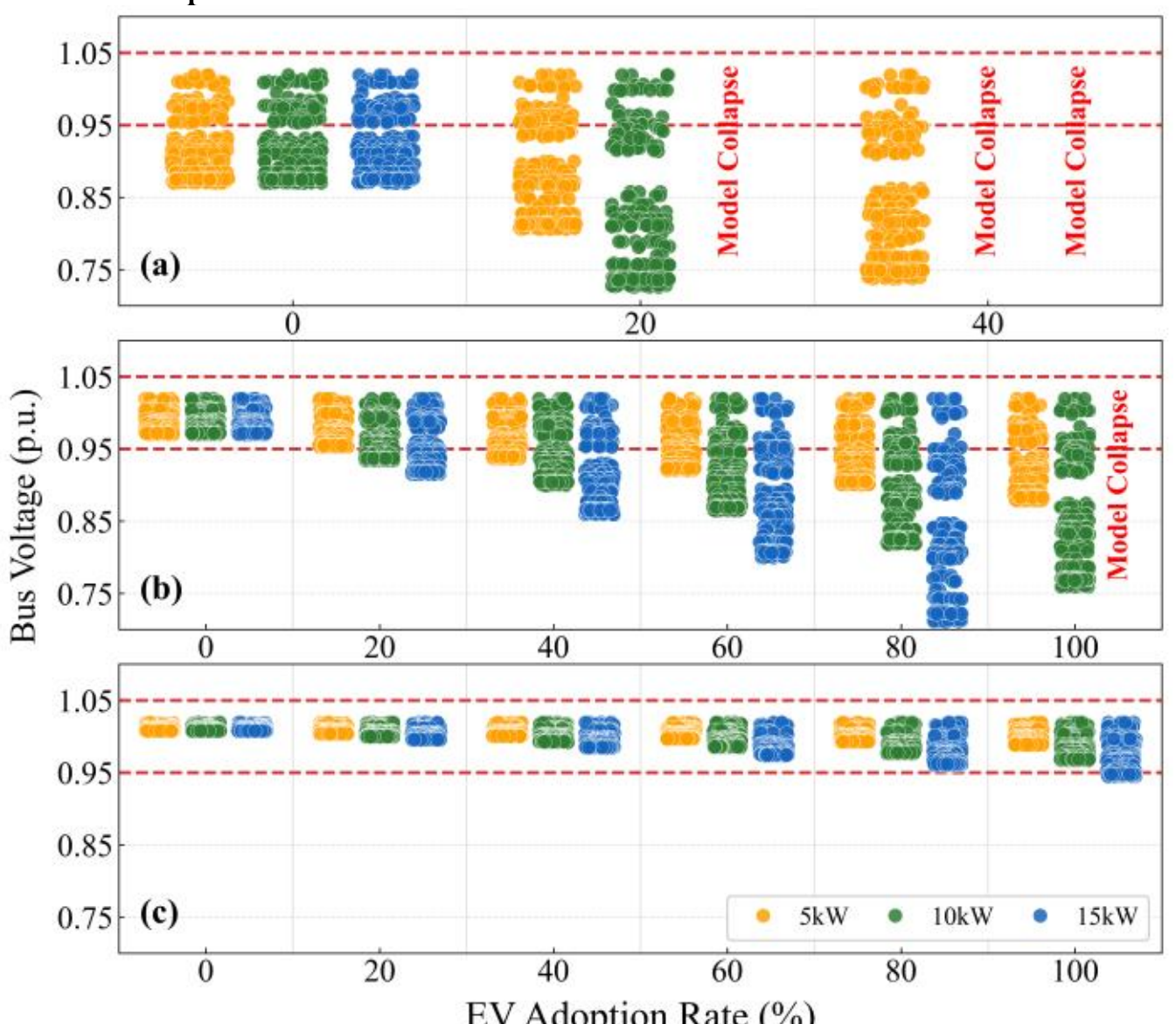

Fig. 7. Bus voltage distribution at: (a) 4.16 kV system; (b) 6.9 kV system; (c) 13.8 kV system.

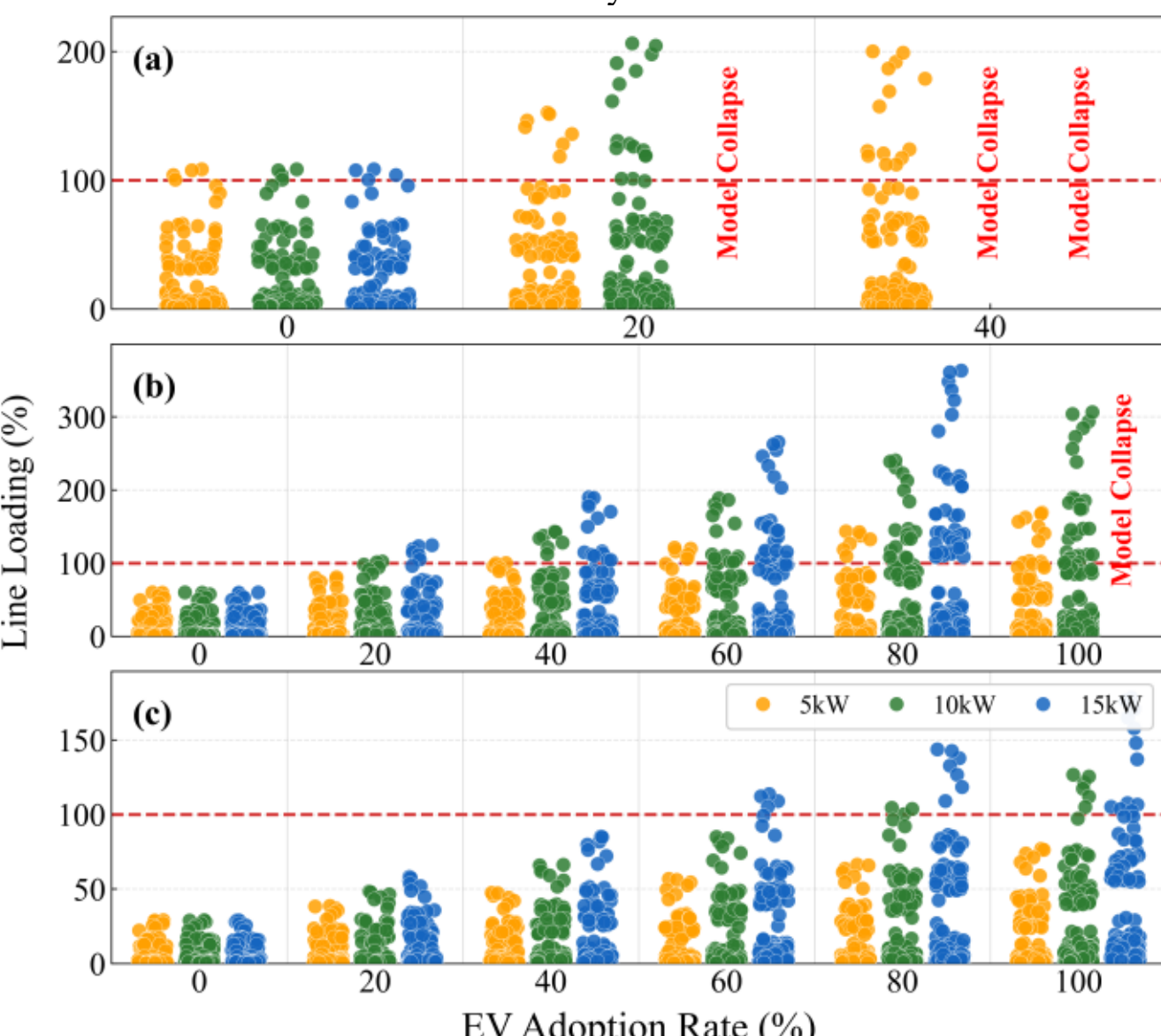

Fig. 8. Line loading percentage distribution at: (a) 4.16 kV system; (b) 6.9 kV system; (c) 13.8 kV system.

Figs. 7 and 8 display scatter plots of bus voltage and line loading percentage corresponding to base voltages of 4.16, 6.9, and 13.8 kV and three different EV charger power levels to compare system performance under varying conditions. Higher EV charger power levels significantly exacerbate grid constraints, resulting in increased violation counts and more severe voltage deviations. Notably, the configuration with 6.9 kV base voltage, 10 kW charging power, and 100% adoption results in concurrent line capacity and voltage violations. This severe scenario serves as the baseline in the following section, where we demonstrate methods to resolve these system criticalities.

## IV. Analysis of Violation Mitigation Approaches

This section presents a comparative analysis of various mitigation approaches and their effectiveness in resolving system violations. Furthermore, this section also evaluates the EV hosting capacity of the distribution network by comparing the baseline system with the BESS-integrated system at different base voltage levels.

### *A. Assessment and Enhancement of EV Hosting Capacity*

Under the 10-kW EV charger scenario, simulation results pinpoint the EV adoption thresholds triggering voltage or thermal violations at 0%, 13%, and 76% for the base voltages of 4.16 kV, 6.9 kV, and 13.8 kV, respectively. Exceeding these benchmarks necessitates corrective interventions, such as BESS deployment, voltage uprating, or line reinforcement. However, the efficacy of BESS will be constrained by the grid's residual capacity to support battery charging. Therefore, this study identifies the secondary thresholds representing the maximum EV adoption rates at which BESS remains a viable mitigation strategy: 9% at 4.16 kV, 32% at 6.9 kV and 90% at 13.8 kV as shown in Fig. 9. BESS installation is recommended specifically within the adoption windows bounded by these initial violation points and the identified feasibility limits. Because the model with a 4.16 kV voltage level collapses under a 100% adoption rate of 10-kW EV charger, no further research on solutions for this voltage level will be conducted.

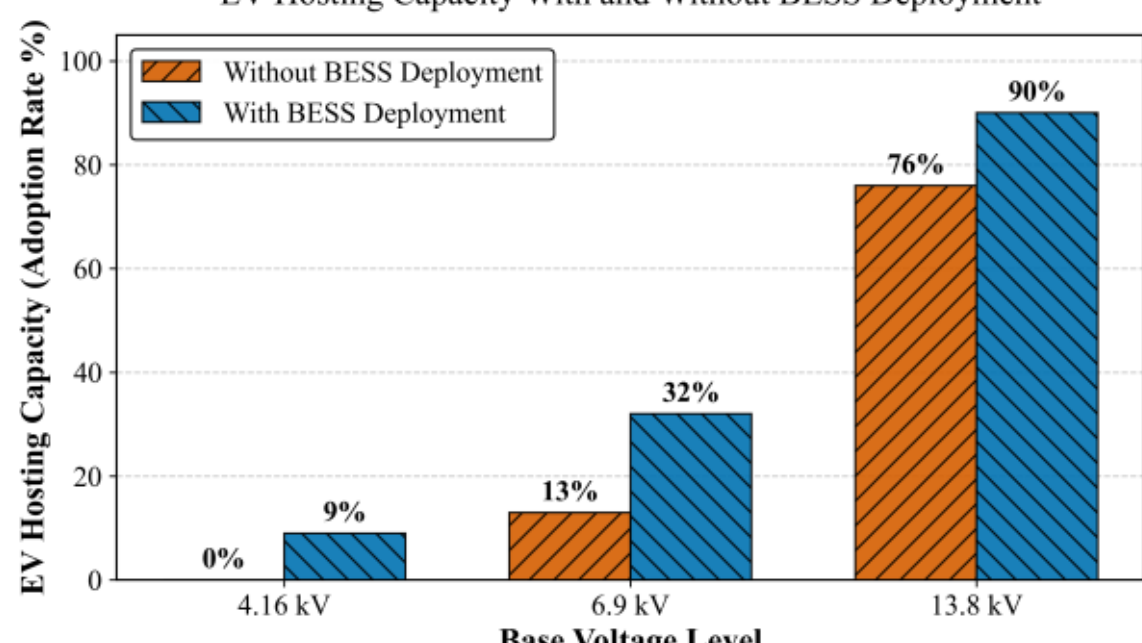

Fig. 9. Comparison of EV hosting capacity with and without BESS deployment across different base voltage levels.

## B. *Results of VCU Model*

Based on the VCU model detailed in Section II.B, the severe thermal and voltage violations observed in the extreme case with 10 kW EV charging power and 100% EV penetration under the 6.9 kV network are successfully mitigated solely by upgrading the distribution cables. The types of the candidate set of upgrade cables and the estimated cost are detailed in Table IV [33], while the fixed one-time cost of circuit breaker upgrades is $18,000 for a 630 A circuit breaker and $24,000 for a 1250 A circuit breaker.

TABLE IV. THE TOTAL CAPITAL COST REQUIRED FOR CABLE UPGRADE

| **Type** | **120mm$^2$** | **150mm$^2$** | **185mm$^2$** | **240mm$^2$** | **300mm$^2$** |
|---|---|---|---|---|---|
| **Price** | $98/m | $115/m | $130/m | $155/m | $180/m |
| **Type** | **400mm$^2$** | **500mm$^2$** | **630mm$^2$** | **800mm$^2$** | **2*500mm$^2$** |
| **Price** | $220/m | $260/m | $310/m | $380/m | $460/m |

As detailed in Table II, the VDQ model identified 30 lines experiencing thermal overloads. The branches from bus 1 to bus 3001, from bus 1 to 2001, from bus 2012 to 2013, and from bus 3075 to bus 3076 are identified as circuit breakers and their upgrade increases line ampacity while maintaining the same impedance. To establish a baseline for cable upgrades, the reference ampacity for each congested line was defined as 1.05 times unconstrained peak current which refers to the maximum branch current of the 30 lines. Consequently, only cables from the candidate set with an ampacity meeting or exceeding this threshold were considered for selection. The optimization objective was formulated to eliminate all system violations while minimizing the total upgrade cost, yielding the results summarized in Table V. The cost of upgrade cable is obtained by multiplying the upgrade price of each type of cable in Table IV by the line length in [33]. The total cost of the VCU model is $1,264,617.

The optimization results indicate that 27 lines required physical upgrades. In Table V, the Upgraded Cable Type column specifies the exact conductor selected for each reinforced branch, while the Impedance Ratio column denotes the proportional change in line impedance following the upgrade. While standard cable upgrades successfully enhance branch ampacity to resolve thermal overloads, mitigating severe voltage violation issues need additional consideration. Specifically, certain branches must be further upgraded to conductors with larger cross-sectional areas to sufficiently reduce line impedance and the associated voltage drops. Although highly effective at maintaining nodal voltage profiles, this over-sizing strategy inherently incurs substantially higher capital costs.

TABLE V. RESULTS OF THE VCU MODEL

| Bottleneck Branches | Unconstrained Peak Current | Upgraded Cable Type | Upgraded Ampacity | Impedance Ratio |
|---|---|---|---|---|
| (1, 3001) | 998.4 | C.B.-1250 | 1250 | 100% |
| (1, 2001) | 453.6 | C.B.-630 | 630 | 100% |
| (2001, 2002) | 453.5 | 240 mm$^2$ | 485 | 31.60% |
| (2002, 2003) | 390.9 | 185 mm$^2$ | 420 | 29.76% |
| (2003, 2004) | 379.2 | 185 mm$^2$ | 420 | 29.76% |
| (2004, 2005) | 379.2 | 185 mm$^2$ | 420 | 29.76% |
| (2005, 2006) | 374.8 | 240 mm$^2$ | 485 | 25.31% |
| (2006, 2010) | 370.9 | 185 mm$^2$ | 420 | 29.76% |
| (2010, 2011) | 368.5 | 800 mm$^2$ | 860 | 16.61% |
| (2011, 2012) | 368.5 | 500 mm$^2$ | 690 | 18.76% |
| (2012, 2013) | 368.5 | C.B.-1250 | 1250 | 100% |
| (3001, 3003) | 998.4 | 2*500 mm$^2$ | 1200 | 11.87% |
| (3003, 3005) | 990.1 | 2*500 mm$^2$ | 1200 | 11.87% |
| (3005, 3008) | 967.1 | 2*500 mm$^2$ | 1200 | 9.50% |
| (3008, 3015) | 926.4 | 2*500 mm$^2$ | 1200 | 9.50% |
| (3015, 3022) | 871.4 | 2*500 mm$^2$ | 1200 | 9.50% |
| (3022, 3030) | 810.1 | 800 mm$^2$ | 860 | 16.61% |
| (3030, 3031) | 643.6 | 500 mm$^2$ | 690 | 18.76% |
| (3031, 3032) | 638.0 | 500 mm$^2$ | 690 | 18.76% |
| (3032, 3033) | 629.4 | 500 mm$^2$ | 690 | 18.76% |
| (3033, 3034) | 620.7 | 500 mm$^2$ | 690 | 18.76% |
| (3034, 3068) | 612.8 | 500 mm$^2$ | 690 | 18.76% |
| (3068, 3075) | 591.2 | 500 mm$^2$ | 690 | 18.76% |
| (3075, 3076) | 591.2 | C.B.-630 | 630 | 100% |
| (3076, 3079) | 500.2 | 300 mm$^2$ | 540 | 22.64% |
| (3079, 3080) | 500.2 | 300 mm$^2$ | 540 | 22.64% |
| (3080, 3082) | 494.5 | 300 mm$^2$ | 540 | 22.64% |

Note: 'C.B.' denotes the circuit breaker.

## C. *Results of DDCP Framework*

The proposed DDCP framework is employed to address the cost and feasibility limitations of the VCU model and VMBP model. 20 critical bottleneck lines that restrict power transfer and prevent BESS charging are identified from the relaxed-VMBP model in Stage I. In Stage II, rather than relying on oversized conductors, these specific bottlenecks are upgraded based strictly on 1.05 times the diagnostic peak currents of 20 critical bottleneck lines as detailed in Table VI. This targeted sizing perfectly resolves severe thermal overloads, allowing the subsequently deployed BESS to efficiently mitigate all remaining voltage violations.

TABLE VI. CABLE UPGRADE RESULTS UNDER THE DDCP FRAMEWORK

| Bottleneck Branches | Diagnostic Peak Current | Upgraded Cable Type | Upgraded Ampacity | Impedance Ratio |
|---|---|---|---|---|
| (1, 3001) | 933.1 | C.B.-1250 | 1250 | 100% |
| (1, 2001) | 393.4 | C.B.-630 | 630 | 100% |
| (2001, 2002) | 393.4 | 185 mm$^2$ | 420 | 37.15% |
| (3001, 3003) | 979.8 | 2*500 mm$^2$ | 1200 | 11.87% |
| (3003, 3005) | 950.2 | 2*500 mm$^2$ | 1200 | 11.87% |
| (3005, 3008) | 801.7 | 800 mm$^2$ | 860 | 16.61% |
| (3008, 3015) | 767.6 | 800 mm$^2$ | 860 | 16.61% |
| (3015, 3022) | 720.9 | 630 mm$^2$ | 780 | 17.55% |
| (3022, 3030) | 677.1 | 630 mm$^2$ | 780 | 17.55% |
| (3030, 3031) | 538.5 | 400 mm$^2$ | 610 | 20.35% |
| (3031, 3032) | 535.0 | 400 mm$^2$ | 610 | 20.35% |
| (3032, 3033) | 530.0 | 400 mm$^2$ | 610 | 20.35% |
| (3033, 3034) | 513.5 | 300 mm$^2$ | 540 | 22.64% |
| (3034, 3068) | 509.2 | 300 mm$^2$ | 540 | 22.64% |
| (3068, 3075) | 486.4 | 300 mm$^2$ | 540 | 22.64% |
| (3075, 3076) | 486.4 | C.B.-630 | 630 | 100% |
| (3076, 3079) | 430.2 | 240 mm$^2$ | 485 | 25.31% |
| (3079, 3080) | 442.5 | 240 mm$^2$ | 485 | 25.31% |
| (3080, 3082) | 440.2 | 240 mm$^2$ | 485 | 25.31% |
| (3082, 3107) | 343.1 | 150 mm$^2$ | 370 | 43.29% |

In Table VI, the candidate cables are sorted based on their topological locations and the severity of current violations. Furthermore, the diagnostic peak currents presented in Table VI are slightly reduced compared to the unconstrained peak current in Table V. This decrease occurs because the deployed BESS provides active power to reduce line loading and reactive power to mitigate voltage drops. However, critical bottleneck lines in both Table V and Table VI, including branches (3001, 3003) and (3003, 3005) as well as the circuit breaker branch, are assigned the same cable type. Following this, Stage III executes the joint optimization of Top-N cable upgrades and BESS deployment. Additionally, this study assumes the capital expenditures (CAPEX) of a 2-hour BESS installation unit is $250/kWh [34].

TABLE VII. OPTIMAL BESS RESULTS OF DDCP FRAMEWORK

| Number of Upgraded Cables | Optimal BESS Buses and Capacities (kWh) | Cable Upgrade Cost | BESS Installation Cost |
|---|---|---|---|
| 18 | 2058: 995.92, 3080: 1429.85<br>3100: 834.69, 3107: 1983.37 | $957,646.67 | $1,310,960.67 |
| 19 | 2058: 995.93 | $983,158.43 | $248,982.75 |
| 20 | 2058: 995.93 | $1,034,369.40 | $248,982.75 |

The number of upgraded cables, N, represents the Top-N most critical cables selected for upgrading. Given the network topology and the distribution of current limit violations, it is practically justified to prioritize these upgrades sequentially, starting from the most severely congested front-end branches. Under this prioritized sequence, the total planning cost exhibits a clear convex trajectory with respect to N, revealing the optimal point. A comparative analysis is then performed across different N values to determine the corresponding optimal BESS sizing, placement, and total installation costs as shown in Table VII. As the number of upgraded cables increases, the required capacity of the installed BESS decreases, but the final solution needs to consider the costs of both upgrading the cables and installing the BESS. Based on the results in Table VII, Fig. 10 illustrates the convex cost trajectory resulting from the trade-off between cable upgrades and BESS integration.

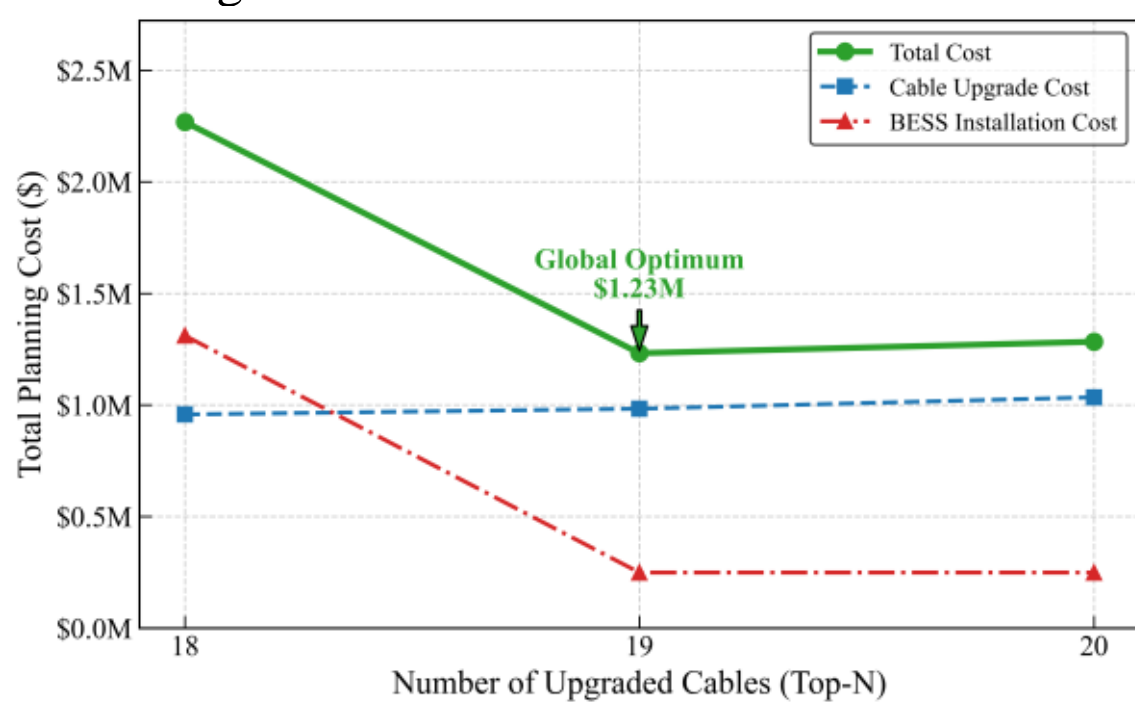

Fig. 10. Optimization of total planning cost: trade-off between cable upgrade cost and BESS installation cost.

As illustrated in Fig. 10 and Table VII, the required BESS capacity stabilizes at approximately 996 kWh when the top 19 or 20 prioritized cables are upgraded. This indicates that extending the upgrades to the top 19 cables effectively eliminates the critical limit violation on branch (3082, 3107). However, restricting the upgrades to only the top 18 cables results in a drastic increase in the required BESS capacity to approximately 5244 kWh. This capacity jump directly corresponds to a severe surge in BESS installation costs. Finally, the most economically viable solution is the co-planning scheme that upgrades the top 19 cables combined with the installation of a 996 kWh BESS at bus 2058. This optimal configuration yields a minimum total investment cost of approximately $1.23 million. Quantitatively, this optimal configuration strictly outperforms the adjacent alternatives by striking a balance between grid infrastructure reinforcement and storage deployment. In the under-investment scenario where only 18 cables are upgraded, the heavy reliance on a massive BESS drives the total planning cost up to $2.27 million. Transitioning to 19 cable upgrades effectively eliminates this operational bottleneck, reducing the total cost to $1.23 million. This represents a substantial saving of $1.04 million, or approximately 45.7%. Furthermore, this optimal point successfully prevents over-investment. Upgrading a 20th cable would unnecessarily inflate the total cost to $1.28 million without yielding further BESS reductions; thus, the 19-cable strategy secures an additional 4.0% saving of roughly $51,200 by avoiding redundant infrastructure expenses.

### *D. Discussion on Voltage Level Upgrade Strategy*

Given that the 240-bus system is capable of operating at a 13.8 kV voltage level, elevating the base voltage presents a viable strategy to mitigate system violations. As demonstrated in Tables II and III, increasing the nominal voltage from 6.9 kV to 13.8 kV eliminates all nodal voltage deviations, leaving only six branches experiencing thermal overloads. Under these improved operating conditions, the remaining network congestion can be fully resolved utilizing the VCU model. The specific upgrade scheme derived from this approach is detailed in Table VIII and the total cost is $250,937.

The feasibility of voltage uprating is heavily dictated by specific operational contexts. It is most practical if higher-voltage insulation standards were proactively adopted during the network's initial design and construction. In such pre-built systems, a voltage upgrade simply involves reconnecting dual-voltage transformers, upgrading surge arresters, and modifying substation configurations. Another viable scenario occurs during end-of-life infrastructure overhauls, where utilities can implement the voltage upgrade concurrently with planned equipment replacements. However, considering the necessary upgrades to some facilities and other expenses after upgrading the voltage level, the required cost will still be higher than the cost of the smaller-scale modifications, such as upgrading cables or installing a BESS.

TABLE VIII. RESULTS OF THE VCU MODEL AT 13.8 KV

| Bottleneck Branches | Unconstrained Peak Current | Upgraded Cable Type | Upgraded Ampacity | Upgrade cost |
|---|---|---|---|---|
| (1, 3001) | 412.67 | C.B.-630 | 630 | $18,000 |
| (3001, 3003) | 412.67 | 240 mm$^2$ | 485 | $108,992 |
| (3003, 3005) | 408.76 | 240 mm$^2$ | 485 | $94,346 |
| (3005, 3008) | 398.62 | 185 mm$^2$ | 420 | $5,389 |
| (3008, 3015) | 380.71 | 185 mm$^2$ | 420 | $12,759 |
| (3015, 3022) | 356.72 | 185 mm$^2$ | 420 | $11,451 |

### *E. Performance of Different Approaches on Violation Issues*

Following the implementation of either the VCU model or the proposed DDCP framework, all network thermal and voltage violations are successfully eliminated. However, only installing the BESS by optimizing the VMBP model cannot resolve all limit violation issues. Furthermore, the voltage level upgrade strategy is more suitable to the distribution network operating below its original design standards making it significantly less viable for resolving localized violations compared to cable upgrades or BESS deployments.

To ensure a fair comparison of the results, all schemes were subjected to the identical 24-hour worst-case scenario. Besides, all violation issues must be completely eliminated, thereby ensuring that every scenario was evaluated against an equal baseline. Because the evaluation addresses extreme worst-case scenarios rather than typical daily EV load profiles, the comparison is explicitly scoped to the initial capital expenditure (CAPEX). While this omits equipment lifespans and BESS arbitrage potential, it guarantees a fair and practically valuable comparative objective. As illustrated in Fig. 11, the hourly nodal voltage profiles for both approaches are strictly maintained above the lower limit of 0.95 per unit.

Fig. 11. Comparison of hourly voltage distribution results between the VCU model and DDCP framework.

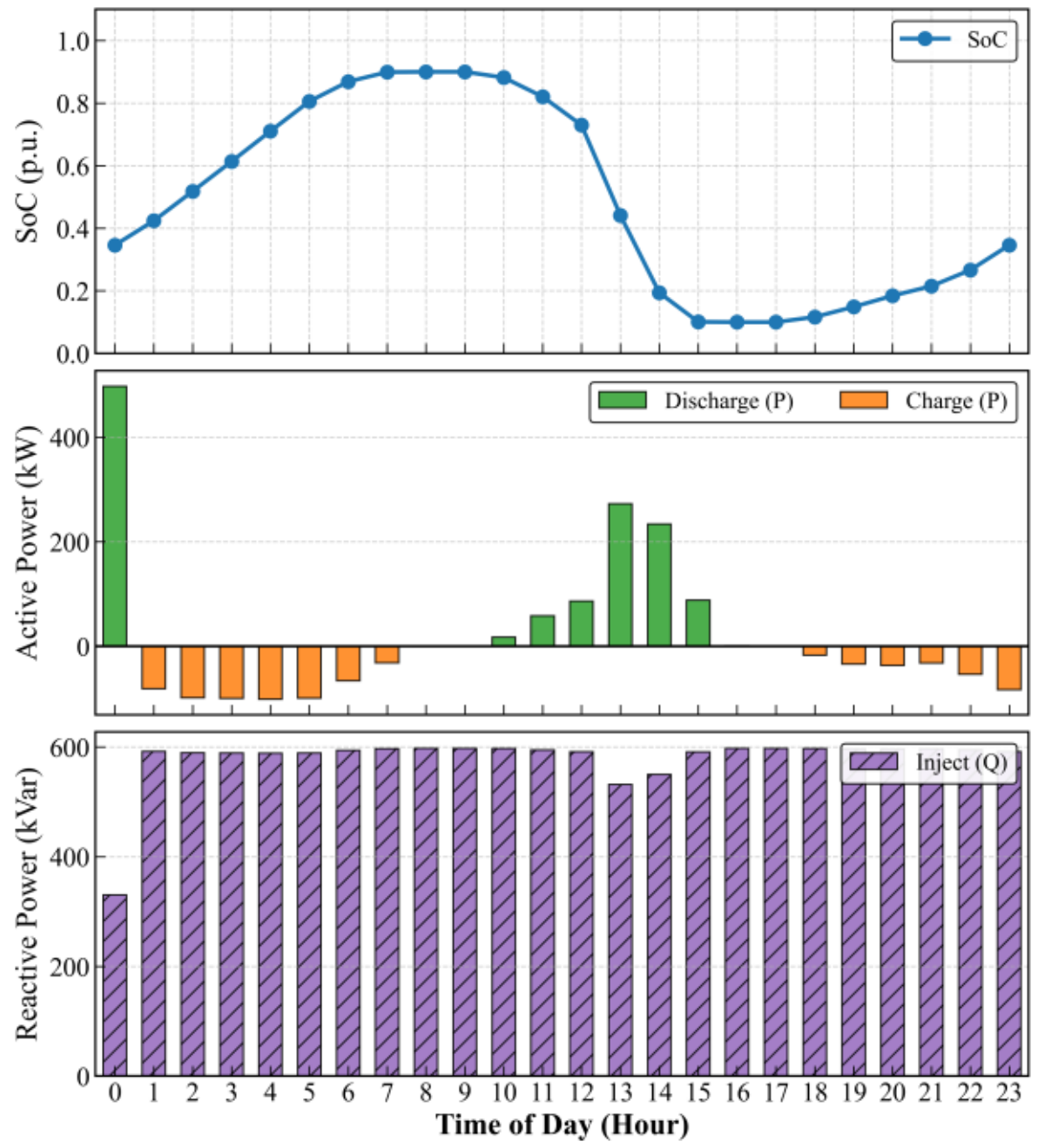

Fig. 12. Diagram of BESS daily operational profile at 6.9 kV.

Fig. 12 illustrates the daily operational profile of the BESS deployed under the DDCP framework. To maintain daily energy balance, the State of Charge (SOC) returns to its initial value at the end of the scheduling horizon. Operationally, the BESS discharges to supply active power to the grid during peak load periods and late-night hours, while it charges during off-peak intervals. Furthermore, the BESS continuously injects reactive power into the distribution network to provide sustained voltage support. While the current DDCP framework already demonstrates lower capital costs than the VCU model, its economic superiority becomes increasingly pronounced over time. Throughout the system's long-term operation, the BESS installed under the DDCP framework can generate substantial operational profits by charging during low-price valleys and discharging during high-price peak periods.

## V. Conclusion

The transition to EVs introduces critical challenges to distribution grid reliability and efficiency. This study comprehensively investigates the impact of varying EV adoption rates and charging capacities on network operational limits. Results demonstrate that higher EV penetrations severely stress distribution networks; for instance, at a 6.9 kV base voltage, violations emerge at just a 13% EV adoption rate, whereas 13.8 kV networks can accommodate up to 76%. Furthermore, 15 kW chargers precipitate system congestion significantly earlier than 5 kW chargers, particularly in lower-voltage systems.

Standard mitigation strategies exhibit fundamental technical, economic, and computational limitations. The VMBP model is restricted by feasibility bounds and completely fails to resolve system violations under extreme scenarios. Similarly, while elevating the system voltage level drastically reduces voltage deviations, it still leaves 6 critical lines requiring physical reinforcement; factoring in the requisite substation and equipment overhauls, the total capital expenditure for voltage uprating remains prohibitively high. Furthermore, directly formulating a joint optimization for simultaneous cable upgrades and BESS installation proves computationally intractable.

To overcome these critical deficiencies, the proposed DDCP framework systematically decouples the problem. By diagnosing bottleneck lines and executing targeted reinforcements prior to BESS allocation, the framework guarantees system compliance while avoiding conservative over-sizing. Economically, the DDCP framework identifies an optimal configuration totaling approximately $1.23 million, successfully outperforming the VCU model, which costs $1.26 million.

## VI. Future Work

While this study provides robust planning insights, it is predicated on a deterministic, simultaneous EV charging profile. Future research will incorporate stochastic charging behaviors by modeling random arrival times and dynamic loads via probabilistic distributions. Exploring demand-side solutions will be essential to further minimize infrastructure investments and maintain grid stability in densely populated EV regions. Furthermore, by incorporating the actual service lifespans of the equipment and the arbitrage potential of the BESS, the optimization model will be able to realize the optimal solution by balancing the arbitrage advantages of the BESS against the risks associated with its shorter service life.

Even though this study focuses on the uncertainties and distributed load surges of EV integration, the proposed DDCP framework can be naturally extended to address the emerging challenges of small-to-medium data centers in distribution networks in the future. Data centers impose severe, continuous stress on cable loading and are often integrated with internal BESS. Formulating a joint optimization model for cable upgrading and BESS configuration inevitably still leads to an intractable computational problem. The DDCP framework serves as an excellent and viable technical pathway to solve these highly complex, concentrated load congestion issues in future grid planning.

While the proposed framework is inherently adaptable to practical unbalanced networks, this study adopted a balanced three-phase approximation to maintain the computational tractability of the large-scale MILP formulation. This provides an efficient and economically viable planning baseline, but it may underestimate the specific voltage drops on heavily loaded phases in severely unbalanced real-world feeders. To extend the framework to three-phase unbalanced feeders, a three-phase linearized branch flow model can be adopted in the optimization layer to obtain a tractable planning formulation, while the resulting investment decisions should be validated using a high-fidelity power flow solver. The decision variables for cable upgrades and BESS installations should be expanded to become phase specific, allowing for targeted capacity enhancements on individual, heavily loaded phases at the cost of increased computational complexity.

# REFERENCES

[1] E. Raffoul and X. Li, "Assessing the Impact of Electric Vehicle Charging on Residential Distribution Grids," *IEEE IAS Annual Meeting*, Taipei, Jun. 2025.

[2] EPA (2021), Fast Facts - U.S. Transportation Sector Greenhouse Gas Emissions 1990 – 2019, EPA, U.S., [Available Online]: https://nepis.epa.gov/Exe/ZyPDF.cgi?Dockey=P10127TU.pdf

[3] Wan, Haoxiang, Linhan Fang, and Xingpeng Li. "Grid Operational Benefit Analysis of Data Center Spatial Flexibility: Congestion Relief, Renewable Energy Curtailment Reduction, and Cost Saving." 2025, arXiv:2511.08759.

[4] BloombergNEF (2024), Electric Vehicle Outlook 2024: Executive Summary, Bloomberg Finance L.P. [Available Online]: https://assets.bbhub.io/professional/sites/24/847354_BNEF_EVO2024_ExecutiveSummary.pdf

[5] IEA (2024), Global EV Outlook 2024, IEA, Paris https://www.iea.org/reports/global-ev-outlook-2024

[6] H. Verma, P. Tripathi and A. M. Naqvi, "A comprehensive review on impacts of electric vehicle charging on the distribution network," *2021 International Conference on Control, Automation, Power and Signal Processing (CAPS)*, Jabalpur, India, 2021, pp. 1-6, doi: 10.1109/CAPS52117.2021.9730630.

[7] M. El-Hendawi, Z. Wang, R. Paranjape, S. Pederson, D. Kozoriz and J. Fick, "Impact evaluation of EV integration into residential power distribution system," *2023 IEEE Canadian Conference on Electrical and Computer Engineering (CCECE)*, Regina, SK, Canada, 2023, pp. 255-260, doi: 10.1109/CCECE58730.2023.10288708.

[8] P. Papadopoulos, S. Skarvelis-Kazakos, I. Grau, B. Awad, L. M. Cipcigan and N. Jenkins, "Impact of residential charging of electric vehicles on distribution networks, a probabilistic approach," *45th International Universities Power Engineering Conference UPEC2010*, Cardiff, UK, 2010, pp. 1-5.

[9] D. Keser and G. Poyrazoglu, "The Impact of Electric Vehicle Charging Stations on Power Distribution Grid by Statistical and Probabilistic Simulation," *2020 2nd Global Power, Energy and Communication Conference (GPECOM)*, Izmir, Turkey, 2020, pp. 143-147, doi: 10.1109/GPECOM49333.2020.9247919.

[10] P. Roy *et al*., "Impact of Electric Vehicle Charging on Power Distribution Systems: A Case Study of the Grid in Western Kentucky," in *IEEE Access*, vol. 11, pp. 49002-49023, 2023, doi: 10.1109/ACCESS.2023.3276928.

[11] S. I. Ahmed, H. Salehfar and D. F. Selvaraj, "Impact of Electric Vehicle Charging on the Performance of Distribution Grid," *2021 IEEE 12th International Symposium on Power Electronics for Distributed Generation Systems (PEDG)*, Chicago, IL, USA, 2021, pp. 1-8.

[12] D. Anggraini, M. Amelin and L. Söder, "Electric Vehicle Charging Considering Grid Limitation in Residential Areas," *2024 IEEE Transportation Electrification Conference and Expo (ITEC)*, Chicago, IL, USA, 2024, pp. 1-6, doi: 10.1109/ITEC60657.2024.10598892.

[13] Fang, Linhan, Jesus Silva-Rodriguez, and Xingpeng Li. "Data-Driven EV Charging Load Profile Estimation and Typical EV Daily Load Dataset Generation," 2025, *arXiv:2511.13861*.

[14] S. Qiu, K. Rajashekara and R. Pottekkat, "Distributed Control of Multi-Voltage-Level DC Microgrids for Resilient Renewable Interconnection," in *IEEE Transactions on Sustainable Energy*, doi: 10.1109/TSTE.2026.3682722.

[15] Bu, Fankun, et al. "A time-series distribution test system based on real utility data." *2019 North American Power Symposium (NAPS)*. IEEE, 2019.

[16] Fatima, Rida, Linhan Fang, and Xingpeng Li. "A Reliability-Cost Optimization Framework for EV and DER Integration in Standard and Reconfigurable Distribution Network Topologies," 2025, arXiv:2511.02250.

[17] Li, Changfu, Vahid R. Disfani, Hamed Valizadeh Haghi, and Jan Kleissl. "Coordination of OLTC and smart inverters for optimal voltage regulation of unbalanced distribution networks." *Electric Power Systems Research*, 187 (2020): 106498.

[18] Z. Guo, W. Wei, Y. Liu, W. Hu and C. Hu, "Real-Time Distribution Network Reconfiguration Using Multi-Stage Robust Optimization," in *IEEE Transactions on Smart Grid*, vol. 16, no. 6, pp. 4475-4487, Nov. 2025, doi: 10.1109/TSG.2025.3589554.

[19] Ranjusha, K. P., and B. Amutha. "Sustainable development of E-mobility in urban areas using knowledge-based artificial network (KANM), behavioral learning theory (BLT) and distributed optimization algorithm (DOA)." *International Journal of Information Technology*, (2026): 1-11.

[20] Jin Lu, Linhan Fang, Fan Jiang, and Xingpeng Li, "A Black Start Strategy for Hydrogen-integrated Renewable Grids with Energy Storage Systems", *IEEE International Conference on Energy Technologies for Future Grids*, Wollongong, Australia, Dec. 2025.

[21] Fang, Linhan, and Xingpeng Li. "Optimal BESS Sizing and Placement for Mitigating EV-Induced Voltage Violations: A Scalable Spatio-Temporal Adaptive Targeting Strategy," 2025, arXiv:2511.00297.

[22] M. Kabirifar, M. Fotuhi-Firuzabad, M. Moeini-Aghtaie, N. Pourghaderi and M. Shahidehpour, "Reliability-Based Expansion Planning Studies of Active Distribution Networks With Multiagents," in *IEEE Transactions on Smart Grid*, vol. 13, no. 6, pp. 4610-4623, Nov. 2022, doi: 10.1109/TSG.2022.3181987.

[23] S. Qiu, K. Rajashekara and R. Pottekkat, "Novel Unified Control Framework for Networked RES-BES Microgrids With Enhanced Performance," in *IEEE Transactions on Sustainable Energy*, vol. 17, no. 1, pp. 30-42, Jan. 2026, doi: 10.1109/TSTE.2025.3597035.

[24] M. Al-Saffar and P. Musilek, "Reinforcement Learning-Based Distributed BESS Management for Mitigating Overvoltage Issues in Systems With High PV Penetration," in *IEEE Transactions on Smart Grid*, vol. 11, no. 4, pp. 2980-2994, July 2020, doi: 10.1109/TSG.2020.2972208.

[25] Hassan Zahid Butt and Xingpeng Li, "Enhancing Optimal Microgrid Planning with Adaptive BESS Degradation Costs and PV Asset Management: An Iterative Post-Optimization Correction Framework", *Electric Power Systems Research*, vol. 247, Oct. 2025.

[26] J. Hu, S. You, M. Lind and J. Østergaard, "Coordinated Charging of Electric Vehicles for Congestion Prevention in the Distribution Grid," in *IEEE Transactions on Smart Grid*, vol. 5, no. 2, pp. 703-711, March 2014, doi: 10.1109/TSG.2013.2279007.

[27] S. Deb *et al*., "Charging Coordination of Plug-In Electric Vehicle for Congestion Management in Distribution System Integrated With Renewable Energy Sources," in *IEEE Transactions* on Industry Applications, vol. 56, no. 5, pp. 5452-5462, Sept.-Oct. 2020, doi: 10.1109/TIA.2020.3010897.

[28] J. Then, A. P. Agalgaonkar and K. M. Muttaqi, "Coordinated Charging of Spatially Distributed Electric Vehicles for Mitigating Voltage Rise and Voltage Unbalance in Modern Distribution Networks," in *IEEE Transactions on Industry Applications*, vol. 59, no. 4, pp. 5149-5157, July-Aug. 2023, doi: 10.1109/TIA.2023.3273186.

[29] Yan, Ziming, Tianyang Zhao, Yan Xu, Leong Hai Koh, Jonathan Go, and Wee Lin Liaw. "Data-driven robust planning of electric vehicle charging infrastructure for urban residential car parks." *IET generation, transmission & distribution*, 14, no. 26 (2020): 6545-6554.

[30] Cooper, C. B. "IEEE recommended practice for electric power distribution for industrial plants." *Electronics and Power*, 33.10 (1987): 658.

[31] Fang, Linhan, and Xingpeng Li. "Violation-Informed Spatio-Temporal Adaptive Targeting Framework for EV-Driven Distribution System Expansion Planning." 2026, arXiv:2606.11638.

[32] T. Saikawa, K. Tanaka and K. Tanaka, "Formal Verification and Code-Generation of Mersenne-Twister Algorithm," *2020 International Symposium on Information Theory and Its Applications* (ISITA), Kapolei, HI, USA, 2020, pp. 607-611.

[33] Cables, Prysmian. "High voltage cables." Toronto, Canada pt I (2015): 80.

[34] Cole, Wesley, A. Will Frazier, and Chad Augustine. Cost projections for utility-scale battery storage: 2021 update. Golden, CO: National Renewable Energy Laboratory, 2021.